\RequirePackage{fix-cm} 
\documentclass[a4paper,reqno,twoside,11pt]{amsart}

\usepackage{graphicx}
\usepackage{a4wide,verbatim}
\usepackage[english]{babel}
\usepackage[latin1]{inputenc}
\usepackage[T1]{fontenc}
\usepackage[dvipsnames]{xcolor}
\usepackage{epstopdf, epsfig}
\usepackage{amsmath,amsfonts,amssymb,amsgen,amsbsy,dsfont}
\usepackage{upgreek}
\usepackage{subfig}
\usepackage{bm}
\usepackage{hyperref}
\usepackage{interval}
\usepackage{cleveref}

\usepackage[square,numbers]{natbib}
\newcommand{\depth}{d}                              
\newcommand{\sur}[1]{{#1}_\mathrm{s}}                    
\renewcommand{\bot}[1]{{#1}_\mathrm{b}}                  
\newcommand{\ave}[1]{\bar{#1}}                      
\newcommand{\eqdef}{\stackrel{\text{\tiny{def}}}{=}} 
\newcommand{\subtag}[1]{\tag{\theparentequation#1}}
\newlength{\intwidth}
\DeclareRobustCommand{\fpint}[2]
   {\mathop{%
      \text{%
        \settowidth{\intwidth}{$\int$}%
        \makebox[0pt][l]{\makebox[\intwidth]{$-$}}%
        $\int_{#1}^{#2}$}}}
        
\DeclareRobustCommand{\ffpint}[2]
   {\mathop{%
      \text{%
        \settowidth{\intwidth}{$\int$}%
        \makebox[0pt][l]{\makebox[\intwidth]{$=$}}%
        $\int_{#1}^{#2}$}}}

\newcommand{\rf}[1]{(\ref{#1})}
\newcommand*\conj[1]{\bar{#1}}

\newcommand{\tr}{\textcolor{red}}
\newcommand{\tb}{\textcolor{blue}}

\newcommand{\ud}{\mathrm{d}}
\newcommand{\uD}{\mathrm{D}}
\newcommand{\ue}{\mathrm{e}}
\newcommand{\ui}{\mathrm{i}}
\newcommand{\upi}{\pi}
\newcommand{\uzeta}{\zeta}

\newcommand{\vu}{\boldsymbol{u}}                    
\newcommand{\vx}{\boldsymbol{x}}                    
\newcommand{\vpsi}{\boldsymbol{\psi}}               
\newcommand{\nab}{\boldsymbol{\nabla}}              
\newcommand{\scal}{\boldsymbol{\cdot}}              

\renewcommand{\Re}{\operatorname{Re}}
\renewcommand{\Im}{\operatorname{Im}}
\newcommand{\plog}[2]{\mbox{\,L{\scriptsize{i}}}_{#1}#2}    
\DeclareMathOperator{\atantwo}{atan2}

\newcommand{\half}{{\textstyle{\frac{1}{2}}}}
\newcommand{\ihalf}{{\textstyle{\frac{\ui}{2}}}}
\newcommand{\halfi}{{\textstyle{\frac{1}{2}\ui}}}
\newcommand{\third}{{\textstyle{\frac{1}{3}}}}
\newcommand{\quat}{{\textstyle{\frac{1}{4}}}}
\newcommand{\fifth}{{\textstyle{\frac{1}{5}}}}
\newcommand{\sixth}{{\textstyle{\frac{1}{6}}}}
\newcommand{\isixth}{{\textstyle{\frac{\ui}{6}}}}
\newcommand{\twothird}{{\textstyle{\frac{2}{3}}}}
\newcommand{\threehalf}{{\textstyle{\frac{3}{2}}}}

\def\Xint#1{\mathchoice
   {\XXint\displaystyle\textstyle{#1}}%
   {\XXint\textstyle\scriptstyle{#1}}%
   {\XXint\scriptstyle\scriptscriptstyle{#1}}%
   {\XXint\scriptscriptstyle\scriptscriptstyle{#1}}%
   \!\int}
\def\XXint#1#2#3{{\setbox0=\hbox{$#1{#2#3}{\int}$}
     \vcenter{\hbox{$#2#3$}}\kern-.5\wd0}}
\def\ddashint{\Xint=}
\def\dashint{\Xint-}

\newcommand{\cs}{c_2}
\newcommand{\ce}{c_1}

\title[General recovery procedure]{General procedure for free-surface recovery from bottom pressure 
measurements: Application to rotational overhanging waves}

\author[J. Labarbe]{Joris Labarbe}
\newcommand{\nfont}{\fontshape{n}\selectfont}
\address{({\nfont\textbf{Joris Labarbe}}) Universit\'e C\^ote d'Azur, CNRS-LJAD UMR 7351,
Laboratoire J. A. Dieudonn\'e, Parc Valrose, F-06108 Nice, France.} 
\email{jlabarbe@unice.fr}

\author[D. Clamond]{Didier Clamond}
\address{({\nfont\textbf{Didier Clamond}}) Universit\'e C\^ote d'Azur, CNRS-LJAD UMR 7351,
Laboratoire J. A. Dieudonn\'e, Parc Valrose, F-06108 Nice, France.} 
\email{didier.clamond@univ-cotedazur.fr}

\subjclass[]{}

\begin{document}
\label{firstpage}
\maketitle

\begin{abstract}
A novel boundary integral approach for the recovery of overhanging 
(or not) rotational (or not) water waves from pressure measurements 
at the bottom is presented. The method is based on the Cauchy 
integral formula and on an  Eulerian--Lagrangian formalism to 
accommodate overturning free surfaces. This approach eliminates the 
need to introduce {\em a priori} a special basis of functions, 
providing thus a general means of fitting the pressure data and, 
consequently, recovering the free surface. The effectiveness and 
accuracy of the method are demonstrated through numerical examples.
\end{abstract}

\maketitle

\section{Introduction}

Nonlinear water waves have been extensively studied since the 
mid eighteenth century, when Euler introduced his eponymous equation.
Since then, surface gravity waves have attracted much attention in 
their modelling, although scientists quickly realised their inherent 
complexity. This is one reason why physicists and mathematicians are 
still interested by the richness of this problem, making it an 
endless source for research in fluid dynamics. For instance, one 
crucial concern in environmental and coastal engineering 
is to accurately measure the surface of the sea for warning about the 
formation of large waves near coasts and oceanic routes. 
One solution to this problem is reconstructing the surface 
using a discrete set of measurements obtained from submerged pressure 
transducers \citep{TT09}. This approach avoids the limitations of offshore buoy systems, 
which are susceptible to climatic disasters, located on moving boundaries, 
and lacking accuracy in wave height estimates \citep{LY20}.
Consequently, solving the nonlinear inverse problem associated with 
water-waves is a timely request for building practical engineering apparatus 
that rely on \textit{in situ} data.

While the hydrostatic theory was originally used to tackle this 
problem when first formulated \citep{BTT69,LW85}, it was only 
recently that nonlinear waves began to be addressed \citep{OVDH}. 
Some works, such as \citep{Constantin2012}, considered conformal 
mapping to successfully obtain reconstruction formulae. However, 
the numerical cost associated with solving these implicit relations 
renders conformal mapping inefficient when dealing with real physical 
data, i.e., when pressure data are given at known abscissas of the 
physical plane, not of the conformally mapped one. Actually, 
introducing some suitable holomorphic functions, it is possible to 
efficiently solve this nonlinear problem  while staying within the 
physical plane for the recovery procedure \citep{ClamondConstantin2013,Clamond2013}.
These studies demonstrated the convergence of the reconstruction 
process and the ability to recover waves of maximum amplitude 
\citep{ClamondHenry2020}. Furthermore, the method was adapted to 
handle cases involving linear shear currents \citep{CLH23}, 
remarkably recovering the unknown magnitude of vorticity alongside 
the wave profile and associated parameters.

The recovery method studied in \citep{Clamond2013,ClamondConstantin2013,
ClamondHenry2020,CLH23} has two shortcomings, however. First, it 
was developed for non-overhanging waves, so it cannot directly 
address these waves that can occur, in particular, in presence of 
vorticity. Second, part of this reconstruction procedure is based 
on analytic continuation of some well-chosen eigenfunctions. If 
for some waves a ``good'' choice of eigenfunctions is clear 
{\em a priori}, this is not necessarily the case for complicated 
wave profiles. By ``good'' choice, we mean a set of eigenfunctions 
that provides an accurate representation of the free surface with 
a minimal amount of modes and, at the same time, that can be easily 
computed. Even though Fourier series (or integrals) can be used in 
principle, a large number of eigenfunctions may be required to 
accurately represent the free surface. Since a surface recovery 
from bottom pressure is intrinsically ill-conditioned, using a 
large number of Fourier modes may lead to numerical issues. Another 
basis should then be preferably employed. For instance, for 
irrotational long waves propagating in shallow water (i.e., 
cnoidal waves), the use of Jacobian elliptic functions is 
effective \citep{Clamond2013,ClamondConstantin2013}. However, 
such alternative basis are not always easily guessed.  
Thus, it is desirable to derive a reconstruction procedure 
independent of a peculiar basis of eigenfunctions. Here, we 
propose a recovery methodology addressing 
these two shortcoming.
 
In this article, we derive a general formulation to address the  
surface recovery problem using a boundary integral formulation. 
While a similar approach was described by \citet{DSP88} for computing 
rotational waves, to our knowledge it has never been applied in the context of a recovery procedure. 
The Cauchy integral formula, although singular by definition, proves 
advantageous from a numerical perspective. 
Dealing with singular kernels, the integral formulation easily allows for the 
consideration of arbitrary steady rotational surface waves (periodic, solitary, 
aperiodic) without the need to select a peculiar basis of functions to fit the 
pressure data. Additionally, this method facilitates the parametrisation of 
the surface profile, enabling the recovery of overhanging waves with arbitrarily  
large amplitudes. 

Overturning profiles are known to be hard to compute accurately \citep{VB94}, 
which presents a significant challenge due to the ill-posed nature of our problem. 
Nevertheless, by considering a mixed Eulerian--Lagrangian description at the 
boundaries, we demonstrate the feasibility of the recovery process. 
To illustrate the robustness of our method, we present two examples of rotational 
steady waves: a periodic wave with an overturning surface and a solitary wave. 
In both scenarii, we achieve good agreement in recovering the wave elevation, 
albeit with the necessity of using refined grids in the regions of greatest 
surface variation. Refining a grid where needed is way much easier than finding a better 
basis of functions, that is a feature of considerable practical interest. 

The method presented in this study consists in the first boundary integral 
approach to solve this nonlinear recovery problem. 
We expect this work to pave the way for addressing even more challenging 
configurations, including the extension to three-dimensional settings, which 
will inevitably involve Green functions in the integral kernels. 

The paper is organised as follow. The mathematical model and relations of 
interest are introduced in section 
\ref{sec1}, with an Eulerian description of motion. In order to handle 
overhanging waves, their Lagrangian counterparts are introduced in section 
\ref{sec2}. In section \ref{sec3}, we derive an equation for the free surface, 
allowing us to compute reference solutions for testing our recovery procedure. 
This procedure is described in the subsequent section \ref{sec4}. Numerical 
implementation and examples are provided in section \ref{sec5}.
Finally, in section \ref{sec6}, we discuss our general results, as well as 
the future extensions and implications of this work.

\section{Mathematical settings}\label{sec1}

We give here the classical formulation of our water-wave problem in 
Eulerian variables. Physical assumptions and notations being identical 
to that of \citet{CLH23}, interested readers should refer to this paper 
for further details. 

\subsection{Equations of motion and useful relations}

\begin{figure}
    \centering
    \includegraphics[width=.8\textwidth]{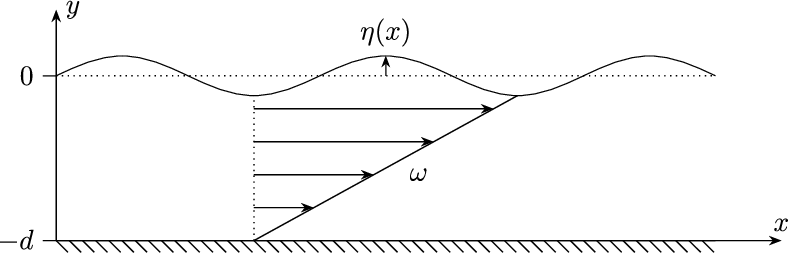}
    \caption{Definition sketch in the referential moving with the wave.}
    \label{Fig1}
\end{figure}

We consider the steady two-dimen\-sional motion of an incompressible inviscid fluid 
with constant vorticity $\omega$. The fluid is bounded above and below by impermeable 
free surface and solid horizontal bed, respectively. Our focus lies on traveling 
waves of permanent form that propagate with a constant phase speed $c$ and wavenumber 
$k$ ($k=0$ for solitary and more general aperiodic waves). We adopt a Galilean frame 
of reference moving with the wave, thus ensuring that the velocity field appears 
independent of time for the observer. Consequently, we can express the fluid domain, 
denoted as $\Omega$, as the set of points $(x,y)$ (Cartesian coordinates) satisfying 
$x \in \mathds{R}$ and $-\depth \leqslant y \leqslant \eta(x)$, where $\eta(x)$ 
represents the surface elevation from rest and $\depth$ is the mean water depth. Thus, 
the mean water level is located at $y=0$, such that
\begin{equation}\label{etamean}
\langle \eta \rangle \eqdef \frac{k}{2\pi} \int_{-\pi/k}^{\pi/k} \eta(x) \ud x = 0 ,
\end{equation}
where $\langle\scal\rangle$ denotes the Eulerian averaging operator (c.f. 
Figure \ref{Fig1}).

In this setting, the velocity field $\bm{u}=(u,v)$ and pressure $p$ 
(divided by the density and relative to the reference value at the surface) 
are governed by the stationary Euler equations
\refstepcounter{equation}
\[
\bm{\nabla} \scal \bm{u}=0, \qquad 
\bm{u} \scal \bm{\nabla} \bm{u} + \bm{\nabla} p + \bm{g} = 0, 
\eqno{(\theequation{\mathit{a},\mathit{b}})}\label{euler}
\]
where $\bm{g}=(0,g)$, the acceleration due to gravity $g>0$ acting downwards. 
Equations of motion \eqref{euler} are supplemented with kinematic and dynamic 
conditions at the upper and lower boundaries
\begin{subequations}\label{bc}
\begin{align}
\bm{u} \scal \bm{n} - \bm{u} \scal \bm{\nabla} \eta &= 0  \quad \textrm{at} \quad y=\eta(x) , \\
p &= 0  \quad \textrm{at} \quad y=\eta(x) , \\
\bm{u} \scal \bm{n} &= 0  \quad \textrm{at} \quad y=-d ,
\end{align}
\end{subequations}
where $\bm{n}$ is the outward normal vector (see Figure \ref{Fig1} 
for a sketch of this configuration).

Since our physical system is two-dimensional \textit{per se}, we introduce a scalar 
stream function $\psi$ such that $u=\psi_y$ and $v=-\psi_x$, so (\ref{euler}{\it a}) 
is satisfied identically and $\omega=-\psi_{xx} -\psi_{yy}$ is assumed constant. Thus, 
the Euler equations can be integrated into the Bernoulli equation 
\begin{equation}\label{Bernoulli}
2 ( p + gy ) + u^2 + v^2 = \sur{B} - 2\omega(\psi-\sur{\psi})\eqdef B(\psi),
\end{equation}
for a constant $\sur{B}$ \citep{CLH23}. Alternatively, we can define a Bernoulli 
constant at the bottom $\bot{B}\eqdef\sur{B} - 2\omega(\bot{\psi}-\sur{\psi})$. 
In \eqref{Bernoulli}, as in the rest of the article, subscripts `s' and `b' denote 
that the fields are evaluated, respectively, at the surface and at the bottom. 
The free surface and the seabed being both streamlines, $\sur{\psi}$ and 
$\bot{\psi}$ are constant in this problem.

Because here $\sur{p}=0$ (constant atmospheric pressure set to zero 
without loss of generality), we have the relations relating some 
average bottom quantities and parameters \citep{CLH23} 
\refstepcounter{equation}
\[
\left<\/\bot{p}\/\right>=g\depth, \qquad 
\left<\/\bot{u}^{\,2}\/\right>=\bot{B},\qquad 
\left<\/\bot{u}-\left(1+\eta_x^{\,2}\right)\sur{u}\/\right>=\omega\/\depth.
\eqno{(\theequation{\mathit{a},\mathit{b},\mathit{c}})}\label{botave}
\]
Although we decide to set the reference frame as moving with the wave celerity, 
it is still useful to consider Stokes' first and second definitions of phase speed
\begin{align}
\ce &\eqdef - \left< \bot{u} \right>
= -\omega\depth - \left< \left(1+\eta_x^{2}\right) \sur{u} \right>, 
\label{defce} \\
\cs &\eqdef - \left< \frac{1}{\depth} \int_{-\depth}^\eta u \ud y \right>
= \frac{\bot{\psi} - \sur{\psi}}{\depth} = - \frac{\omega\depth}{2} - 
\frac{\omega\left<\eta^2\right>}{2\depth} - \frac{\left< \left(1+\eta_x^{2}\right) 
h\, \sur{u} \right>}{\depth},\label{defcs}
\end{align}
where $h\eqdef\eta(x)+\depth$ is the local water depth.

\subsection{Holomorphic functions}

The vorticity being constant, potential theory still holds when using a Helmholtz 
representation to subtract the contribution of the linear shear current \citep{CLH23}. 
Therefore, it is convenient to introduce the complex relative velocity
\begin{equation}\label{W}
W(z) \eqdef U(x,y) - \ui V(x,y) = (y+\depth)\omega  + u(x,y) - \ui v(x,y),
\end{equation}
that is holomorphic in $\Omega$ for the complex coordinate $z\eqdef x+\ui y$. 
(Obviously, the complex velocity $w\eqdef u-\ui v$ is not holomorphic if $\omega\neq0$.)
The relative complex velocity \eqref{W} is related to the relative complex potential 
$F(z)\eqdef \Phi(x,y) + \ui\Psi(x,y)$ by $W=\ud F/\ud z$, where $U=\Phi_x=\Psi_y$ and 
$V=\Phi_y=-\Psi_x$ (see \citet{CLH23} for more details).

Following \citet{ClamondConstantin2013}, we introduce a {complex ``pressure''} 
function as
\begin{align}
\label{P}
{\mathfrak P}(z)\/ \eqdef\/g\depth + \half\sur{B} + \omega (\sur{\psi} - \bot{\psi}) - \half W(z)^2 = g\depth + \half\bot{B}  - \half W(z)^2 , 
\end{align}
that is holomorphic in the fluid domain $\Omega$, its restriction to the flat bed 
$y=-\depth$ having zero imaginary part and real part $p_\text{b}$. 
Thus, $p_\text{b}$ determines $\mathfrak{P}$ uniquely throughout the fluid domain, 
i.e., ${\mathfrak P}(z)=\bot{p}(z+\ui\depth)$. 
Note that $p$ introduced in \eqref{euler} coincides with the real part of 
$\mathfrak{P}$ only on $y=-\depth$ because the former is not a harmonic function 
in the fluid domain \citep{Constantin2006,ConstantinStrauss2010}.  

As for irrotational waves, it is useful \citep{Clamond2013,Clamond2018,CLH23} 
to introduce the anti-derivative of $\mathfrak{P}(z)$
\begin{align} 
\label{defQfun}
\mathfrak{Q}(z) \eqdef \int_{z_0}^z \left[ \mathfrak{P}(z') - g\depth \right]
\ud z' =\int_{z_0}^z \half \left[ \bot{B} - W(z')^2 \right] \ud z', 
\end{align}
where $z_0$ is an arbitrary constant. 
For the same abscissa $x$, the functions $\mathfrak{Q}$ at the free surface 
(i.e., $\sur{\mathfrak{Q}}(x)$) and at the bottom (i.e., $\bot{\mathfrak{Q}}(x)$) 
satisfy the relation
\begin{equation}
\label{QsQbdef}
\sur{\mathfrak{Q}}(x) - \bot{\mathfrak{Q}}(x) = \int_{x-\ui d}^{x+\ui\eta(x)} 
\left[\/ \mathfrak{P}(z)- g\depth\/ \right] \ud z  
= \frac{\ui h(x) \bot{B}}{2} - \int_{x-\ui d}^{x+\ui\eta(x)} \frac{W(z)^2}{2} \ud z.
\end{equation}

\subsection{Cauchy integral formula}

In the complex $z$-plane, the boundaries are analytical curves defined 
by $\sur{z}=x+\ui\eta$ and $\bot{z}=x-\ui d$. For a holomorphic 
function $\Upxi(z)$, the Cauchy integral formula applied to the fluid 
domain $\Omega$ (assuming non-intersecting and non-overturning 
seabed and free surface) yields
\begin{equation}
\ui\vartheta\,\Upxi(z) = \text{P.V.} \oint \frac{\Upxi(z')}{z'-z} \ud z' 
= \int_{-\infty}^\infty \frac{\bot{\Upxi}' \ud x'}{\bot{z}'-z} - 
\int_{-\infty}^\infty \frac{\left(1 + \ui\eta_x'\right) \sur{\Upxi}' \ud x'}
{\sur{z}'-z} ,
\end{equation}
where $\vartheta=\{2\pi,0,\pi\}$ respectively inside, outside and at 
the smooth boundary of the domain. We emphasis that, in this paper, 
all integrals must be taken in the sense of the Cauchy principal value 
(P.V.), even if it is not explicitly mentioned for brevity. When 
$\Im\{\bot{\Upxi}\}=0$, the bottom boundary condition can be taken 
into account with the method of images, yielding 
\begin{equation}\label{cauchyinf}
\Upxi(z) = \frac{\ui}{\vartheta} \int_{-\infty}^{\infty} \frac{ 
\left(1+\ui\eta_x'\right) \sur{\Upxi}'\ud x' }{ \sur{z}' - z } - 
\frac{\ui}{\vartheta} \int_{-\infty}^{\infty} \frac{ \left(1-
\ui\eta_x'\right)\sur{\conj{\Upxi}}' \ud x' }{ \sur{\conj{z}}' - 
z - 2\ui d},
\end{equation}
where an overbar denotes the complex conjugation.
Note that the formula \eqref{cauchyinf} is valid in finite depth 
(provided that $\Im\{\bot{\Upxi}\}=0$), and in infinite depth if 
$\bot{\Upxi}\to0$ as $d\to\infty$. Examples of functions satisfying 
these conditions are $\Upxi=W+c_1$, $\Upxi=W^2-\bot{B}$ and $\Upxi=
\mathfrak{P}-gd$, so \eqref{cauchyinf} provides an expression for 
computing these functions in arbitrary depth.

\subsection{Integral formulations for periodic waves}

For $L$-periodic waves (with $L=2\pi/k$), the kernel is repeated in 
the horizontal direction, along the interval $\mathcal{I}=[0,L]$, 
leading to the Cauchy integral formula with Hilbert kernel
\begin{align}
\label{cauchyperegbot}
\Upxi(z) = \frac{\ui k}{2\vartheta} \int_{\mathcal{I}} \left[ \cot\!\left( 
k\frac{\sur{z}'-z}{2} \right) \left(1+\ui\eta_x'\right) \sur{\Upxi}' 
- \cot\!\left( k\frac{\bot{z}'-z}{2} \right)\bot{\Upxi}' \right] \ud x' . 
\end{align}

Alternatively, using the method of images, along with the identity 
\rf{Licot}, the Cauchy integral can be written 
\begin{align}
\label{cauchypereg}
\Upxi(z) = \frac{k}{\vartheta} \int_{\mathcal{I}} &\left[ \plog{0} \{\ue^{\ui 
k (\sur{z}'-z)}\} \left(1+\ui\eta_x'\right) \sur{\Upxi}' + \plog{0} 
\{\ue^{\ui k (z-\sur{\conj{z}}'+2\ui\depth)}\} \left(1-\ui\eta_x'
\right) \sur{\conj{\Upxi}} \right] \ud x' \nonumber \\
&+ \pi\/\vartheta^{-1} \langle\/ (1+\ui\eta_x)\sur{\Upxi} + 
(1-\ui\eta_x)\sur{\conj{\Upxi}} \rangle ,
\end{align}
where $\plog{\nu}$ is the $\nu$th polylogarithm whose definition 
is given in appendix \ref{appA} along with useful relations.
It is worth noticing that the last term in the right-hand side of 
\eqref{cauchypereg} corresponds to the zeroth Fourier coefficient 
(not present when the holomorphic function $\Upxi(z)$ has zero mean 
over the wave period).
Equation \eqref{cauchypereg} can be rewritten using the identity 
\rf{Lidz}, yielding 
\begin{align}
\label{dcauchypereg}
\Upxi(z) = \frac{\ui}{\vartheta} \int_{\mathcal{I}} \frac{\partial}
{\partial\/z} &\left[ \plog{1} \{\ue^{\ui k (\sur{z}'-z)}\} \left(
1+\ui\eta_x'\right) \sur{\Upxi}' - \plog{1} \{\ue^{\ui k (z-\sur{
\bar{z}}'+2\ui\depth)}\} \left(1-\ui\eta_x'\right) \sur{\bar{\Upxi}} 
\right] \ud\/x' \nonumber\\
&+ 2\/\pi\/\vartheta^{-1} \left< \Re{\left\{\sur{\Upxi}\right\}} - 
\eta_x \Im{\left\{\sur{\Upxi}\right\}} \right> , 
\end{align}

At the free surface (where $\vartheta=\pi$, $z=\sur{z}$ and $\ud 
\sur{z} = (1+\ui\eta_x) \ud\/x$), carefully applying the Leibniz 
integral rule (c.f. formula \rf{relcauchdev2} in appendix \ref{appB})
on the singular term in the integrand, equation \eqref{dcauchypereg} 
reduces to
\begin{align}
\label{dcauchyperegim}
\left(1+\ui\eta_x\right)\sur{\Upxi} = &\frac{\ui}{2\pi} \frac{\ud}{\ud\/x} \int_{\mathcal{I}} \left[ \plog{1} \{ \ue^{\ui k (\sur{z}'-\sur{z})} \} \left(1+\ui\eta_x'\right) \sur{\Upxi}'-\plog{1} \{ \ue^{\ui k (\sur{z}-\sur{\bar{z}}'+2\ui\depth)} \} \left(1-\ui\eta_x'\right) \sur{\bar{\Upxi}} \right] \ud\/x' \nonumber \\
&+ \left(1+\ui\eta_x\right) \left< \Re{\left\{\sur{\Upxi}\right\}} - 
\eta_x \Im{\left\{\sur{\Upxi}\right\}} \right> .
\end{align}

It should be noted that, obviously, aperiodic equations can be obtained 
from the periodic ones letting $L\to\infty$ (i.e. $k\to0^+$). Thus, for 
now on, we only consider periodic waves.

\section{Lagrangian description}\label{sec3}


This section focuses on addressing the limitation of the Eulerian 
framework that hinders the computation of overhanging waves, which 
are characterised by multi-valued surfaces. 
While one option to address this challenge involves employing an 
arclength formulation, as elaborated by \citet{VB94}, we have 
chosen to adopt a Lagrangian formalism in this study. 
One benefit of the Lagrangian approach is its inherent capability 
to aggregate collocation points near the wave crest, where they 
are most crucial \citep{DSP88}.

Our approach is similar to that used by \citet{DSP88}, although we 
modify the integral kernels in terms of polylogarithms and we 
primitive the Cauchy integral formula to remove the strong 
singularity from the kernels following \citet{Clamond2018}.
The resulting analytical expression is characterised by a weak 
logarithmic singularity, and it is suitable for calculating various 
types of waves, including solitary and periodic waves, overhanging 
or not.


Considering the (rotational) complex velocity $w=u-\ui v$ 
evaluated at the surface, let introduce the holomorphic function 
$\log{(-{w}/\sqrt{gd})}={q}-\ui{\theta}$ and define accordingly
\begin{align}
\sur{q} &\eqdef \Re{\{\log{(-\sur{w}/\sqrt{gd})}\}} = \frac{1}{2} 
\ln{ \left[ (\sur{u}^2+\sur{v}^2)/\sqrt{gd} \right] } , \nonumber \\ 
\sur{\theta} &\eqdef -\Im{\{\log{(-\sur{w}/\sqrt{gd})}\}} = 
-\atantwo{(\sur{v},-\sur{u})}, 
\label{theta}
\end{align}
where we notably used the Bernoulli's principle at the surface.

Exploiting the impermeability and isobarity of the free surface, 
one gets 
\begin{equation}
\tan{\sur{\theta}} = \eta_x = \frac{\sur{v}}{\sur{u}} = \frac{\sigma\sur{v}}{\sqrt{\sur{B} - 2g\eta - \sur{v}^2}} = \frac{\sqrt{\sur{B} - 2g\eta - \sur{u}^2}}{\sigma\sur{u}} .
\end{equation}
Hence, extracting $\sur{u}$ and $\sur{v}$ from the latter relations, we have
\begin{equation}
\label{usvs}
\sur{u} = \sigma\cos({\sur{\theta}}) \sqrt{\sur{B} - 2g\eta} , \quad 
\sur{v} = \sigma\sin({\sur{\theta}}) \sqrt{\sur{B} - 2g\eta} .
\end{equation}

We now consider the Lagrangian description of motion, with $t$ denoting 
the time, thus $\sur{u}=\ud x/\ud t$ and $\sur{v}=\ud\eta/\ud t$ ($\ud/
\ud t$ the temporal derivative following the motion). From the second  
expression in \eqref{usvs}, we deduce that
\begin{equation}
\frac{\ud}{\ud t} \sqrt{\sur{B} - 2g\eta} = - \sigma 
g \sin({\sur{\theta}}),
\end{equation}
and hence, considering a crest at $t=0$ (where $\eta(0)=a$ is the wave 
amplitude), we have
\begin{equation}
\sqrt{\sur{B} - 2g\eta} = \mu - g \sigma \int_{0}^{t} 
\sin({\sur{\theta}'})\, \ud t', \quad
\mu \eqdef \sqrt{\sur{B} - 2ga},
\end{equation}
where $\sur{\theta}'\eqdef\sur{\theta}(t')$. Therefore, all 
quantities at the free surface can be expressed in terms 
of the surface angle $\sur{\theta}$, using $t$ as independent variable, 
e.g.
\begin{align}
\sur{z}(t) &= \sigma \int_{0}^{t} \left[ \mu - g \sigma \int_{0}^{t'} 
\sin({\sur{\theta}''})\, \ud t'' \right] \exp{(\ui\sur{\theta}')} 
\ud t' + \ui a , \label{zt} \\
\sur{w}(t) &= \frac{\ud\sur{\bar{z}}}{\ud t} = \sigma \left[ \mu - 
g \sigma \int_{0}^{t} \sin(\sur{\theta}')\,\ud t' \right] \exp{(-\ui\sur{\theta})} , \\
\eta(t) &= \Im{\sur{z}} = \frac{\sur{B}}{2g} - \frac{1}{2g} 
\left[ \mu - g \sigma \int_{0}^{t} \sin(\sur{\theta}')\,\ud t' 
\right]^2 . \label{eta}
\end{align}

In the Eulerian description of motion, the wave period is constant and depends 
on the reference frame (moving at a constant phase speed) where one observes 
the fluid. On the other hand, in the Lagrangian context, the period $T_L$ of 
the free surface differs from the Eulerian period due to the Stokes drift 
\citep{LH2013}. The Lagrangian period being such that $\eta(t+T_L)=\eta(t)$, 
we exploit expression \eqref{eta} which, after some elementary 
algebra and simplifications, yields
\begin{equation}
\left[ 2\mu - g \sigma \int_{t}^{t+T_L} \sin(\sur{\theta}')\, 
\ud t' - 2g\sigma \int_{0}^{t} \sin(\sur{\theta}')\, \ud t' 
\right] \int_{t}^{t+T_L} \sin(\sur{\theta}')\, \ud t' = 0,
\end{equation}
that is necessarily satisfied for all times if and only if
\begin{equation}\label{TL}
\int_{0}^{T_L} \sin(\sur{\theta}')\, \ud t' = 0,
\end{equation}
thus defining the Lagrangian period $T_L$.


\section{Equations for the free surface}\label{sec2}

Since we do not have access to measurements of the bottom pressure 
for rotational waves, we must generate these data from numerical 
solutions of the exact equations. Thus, this section aims to derive 
a comprehensive formulation for the computation of surface wave using 
a boundary integral method. From these solutions, the bottom pressure 
is subsequently obtained and used as input for our surface recovery 
procedure.

\subsection{Eulerian formulation}

From expressions \eqref{Bernoulli} and \eqref{W}, the complex 
(irrotational part of the) velocity at the surface is given 
explicitly by 
\begin{equation}\label{Ws}
\sur{W} = \omega h + \sigma \left(1-\ui\eta_x\right) 
\sqrt{(\sur{B}-2g\eta)/(1+\eta_x^2)},
\end{equation}
$\sigma=\mp 1$ denoting waves propagating upstream or downstream, 
respectively. The parameter $\sigma$ is introduced for convenience 
in order to characterise the (arbitrarily chosen) direction of the 
wave propagation in a `fixed' frame of reference, i.e., $\sigma=-1$ 
if the wave travels toward the increasing $x$-direction in this frame 
and, obviously, $\sigma=+1$ if the wave travels toward the decreasing 
$x$-direction.
   
Considering the holomorphic function $\Upxi=W+c$ ($c$ being an arbitrary 
definition of the phase speed), the left-hand side of \rf{dcauchyperegim} 
follows directly from \eqref{Ws}
\begin{equation}
\label{holos}
\left(1+\ui\eta_x\right)\sur{\Upxi} = \left(\omega h + c\right) \left(1+
\ui\eta_x\right) + \sigma \sqrt{(\sur{B}-2g\eta)(1+\eta_x^2)},
\end{equation}
where the radicand is purely real since $\sur{B}\geqslant\max(2g\eta)$ 
for all waves.
Substituting expression \eqref{holos} in \eqref{dcauchyperegim}, 
the integral term splits into several contributions as
\begin{align}\label{holos2}
\omega \eta \left(1+\ui\eta_x\right) = &\frac{\ui}{2\pi} \frac{\ud}{\ud\/x} 
\left[ \int_{\mathcal{C}} \plog{1} \{ \ue^{\ui k (\sur{z}'-\sur{z})} \} 
\left(\omega h' + c\right) \ud\sur{z}' + \int_{\mathcal{C}} \plog{1} \{ 
\ue^{\ui k (\sur{z}-\sur{\bar{z}}'+2\ui\depth)} \} \left(\omega h' + 
c\right) \ud \sur{\bar{z}}' \right. \nonumber \\
&+\left.\sigma\int_{\mathcal{I}} \plog{\nu}\!\left(\ue^{\ui k(\sur{z}'-
\sur{z})}\right) - \plog{\nu}\!\left(\ue^{\ui k(\sur{z}-
\sur{\bar{z}}'+2\ui\depth)}\right) \sqrt{(\sur{B}-2g\eta)(1+\eta_x^2)} 
\ud x' \right] \nonumber \\
&+ \left(1+\ui\eta_x\right) \langle \sur{u} \left(1+\eta_x^2\right) 
\rangle - \sigma  \sqrt{(\sur{B}-2g\eta)(1+\eta_x^2)} ,
\end{align}
where $\mathcal{C}$ represents the free surface path, i.e., we use 
the brief notations $\int_{\mathcal{C}}(\cdots)\/\ud\sur{z}'\eqdef
\int_{\mathcal{I}}(\cdots)(1+\ui\eta_x')\/\ud\/x'$ and 
$\int_{\mathcal{I}}(\cdots)\/\ud\/x'\eqdef\int_0^L(\cdots)\/\ud\/x'$.
The first term inside the square bracket of the right-hand side of 
\eqref{holos2} reduces to
\begin{align}
J_1 &\eqdef \int_{\mathcal{C}} \plog{1}\!\left(\ue^{\ui k(\sur{z}'-
\sur{z})} \right) (\omega h' + c)\, \ud\/\sur{z}' = \omega 
\int_{\mathcal{C}} \plog{1}\!\left(\ue^{\ui k(\sur{z}'-
\sur{z})}\right) \eta'\,\ud\/\sur{z}'\nonumber \\
&= \omega \int_{\mathcal{I}} \plog{1}\!\left(\ue^{\ui k(\sur{z}'-
\sur{z})} \right) \eta' \left(1+\ui\eta_x'\right) \ud\/x' 
= \frac{\ui\omega}{k} \int_{\mathcal{I}} \plog{2}\!\left(\ue^{\ui 
k(\sur{z}'-\sur{z})}\right)\eta_x'\,\ud\/x' \nonumber \\
&= - \frac{\omega}{k} \int_{\mathcal{I}} \plog{2}\!\left(\ue^{\ui 
k(\sur{z}'-\sur{z})} \right) \ud\/x', \nonumber
\end{align}
where we exploited the property \citep{Clamond2018}
\begin{equation}\label{intplog}
\int_{\mathcal{C}} \plog{\nu}\!\left( \ue^{\ui k\sur{z}} \right) 
\ud \sur{z} = 0.
\end{equation} 
Similarly, we have
\begin{equation}
J_2 \eqdef \int_{\mathcal{C}} \plog{1}\!\left(\ue^{\ui k(\sur{z}-
\sur{\bar{z}}'+2\ui\depth)} \right) (\omega h'+c)\,\ud\/\sur{\bar{z}}' 
= -\frac{\omega}{k} \int_{\mathcal{I}} \plog{2}\!\left(\ue^{\ui 
k(\sur{z}-\sur{\bar{z}}'+2\ui\depth)} \right) \ud\/x'. 
\end{equation}
Finally, after integrating the whole expression \rf{holos2} and retaining 
the imaginary part only, we obtain the equation for the free surface 
\begin{align}
K=&\,\frac{\omega\/\eta^2}{2} - \eta \left< \sur{u} (1+\eta_x^2) \right>  
+ \frac{\omega}{2\pi k} \int_{\mathcal{I}} \Re\{\mathcal{L}_{2}\}\, \ud\/x'
\nonumber\\
&\,- \frac{\sigma}{2\pi} \int_{\mathcal{I}} \Re\{\mathcal{L}_{1}\} \sqrt{(
\sur{B}-2g\eta') (1+\eta_x'^2)}\, \ud\/x',\label{eqeta}
\end{align}
where $K$ is an integration constant 
obtained enforcing the 
mean-level condition \eqref{etamean}, i.e,
\begin{align}
\label{Kcte}
K = \left<\frac{\omega\eta^2}{2} + \frac{\omega}{2\pi k} \int_{\mathcal{I}} \Re\{\mathcal{L}_{2}\}\, \ud\/x'\,- \frac{\sigma}{2\pi} \int_{\mathcal{I}} \Re\{\mathcal{L}_{1}\} \sqrt{(
\sur{B}-2g\eta') (1+\eta_x'^2)}\, \ud\/x'\right>.
\end{align} 
From now on, the same notation $K$ is used to denote 
integration constants in the different surface recovery formulas.
The exact values of these constants is obtained, as 
above, enforcing the condition \eqref{etamean}.
Moreover, we have introduced, for brevity, the notation for the 
kernels
\begin{equation}
\mathcal{L}_{\nu} \eqdef \plog{\nu}\!\left(\ue^{\ui k(\sur{z}'-
\sur{z})}\right) - \plog{\nu}\!\left(\ue^{\ui k(\sur{z}-
\sur{\bar{z}}'+2\ui\depth)}\right).
\end{equation}

Equation \eqref{eqeta} is a nonlinear integro-differential equation for the 
computation of the surface elevation $\eta$. Once $\eta$ is obtained solving 
\eqref{eqeta} numerically, one can compute the corresponding bottom pressure. 
This bottom pressure can then be considered as ``experimental data'' to 
illustrate the reconstruction procedure described below. This is necessary 
when such data are not available from physical measurement, as it is the case 
with rotational waves.

\subsection{Lagrangian formulation}

Expression \rf{TL} provides an implicit definition of $T_L$ 
if the wavelength $L$ is fixed \textit{a priori}. 
It is now of interest to convert our formula for the surface 
elevation \rf{eqeta} using the Lagrangian description 
introduced previously (the variable of interest now becomes 
$\sur{\theta}$). After some simplifications, we obtain
\begin{align}
\frac{\omega\sigma}{2} &\eta^2 - \frac{k\eta}{2\pi} \int_{0}^{T_L} (\sur{B} - 2g\eta) \ud t - \frac{1}{2\pi} \int_{0}^{T_L} \Re{\{\mathcal{L}_1\}} (\sur{B} - 2g\eta') \ud t' \nonumber \\
&+ \frac{\omega\sigma}{2\pi k} \int_{0}^{T_L} \Re{\{\mathcal{L}_2\}} 
\cos\!\left({\sur{\theta}'}\right) \sqrt{\sur{B} - 2g\eta'}\, \ud t' = K ,
\label{eqetaL}
\end{align}
where $K$ is recovered using condition \eqref{etamean} in the same way as to obtain \rf{Kcte}.

The computation of \rf{eqetaL} involves weak logarithmic singularities in the kernel of 
the $\mathcal{L}_1$ operator.
In the numerical implementation, we use a similar approach as 
the one presented by \citet{Clamond2018},  
by subtracting the regular part of the operator. Thence, we obtain an explicit expression 
for the regularized finite integral as
\begin{align}
\label{intL1}
\int_{0}^{T_L} \Re{\{\mathcal{L}_1\}} (\sur{B} - 2g\eta') \ud t' = & -2g 
\int_{0}^{T_L} \Re{\{\mathcal{L}_1\}} (\eta' - \eta) \ud t' \nonumber \\
&+ (\sur{B} - 2g\eta) \int_{0}^{T_L} \Re{\{\mathcal{L}_1 - 
\widehat{\mathcal{L}}_1\}} \ud t',
\end{align}
where we introduced
\begin{equation}
\widehat{\mathcal{L}}_{\nu} \eqdef \plog{\nu} \{ \ue^{\ui \tau(t'-t)}\} - \plog{\nu}\{ \ue^{\ui \tau(t-t')} \ue^{-2k\depth} \},
\end{equation}
with $\tau\eqdef2\pi/T_L$.
Considering $t\to t'$ in both integrands in \rf{intL1}, we have
\begin{align}
\lim_{t\to t'} \Re{\{\mathcal{L}_1\}} (\eta - \eta') &= 0 ,  \\
\lim_{t\to t'} \Re{\{\mathcal{L}_1 - \widehat{\mathcal{L}}_1\}} &= \log\! 
\left[ \left( \frac{1 - \ue^{-2kh}}{1 - \ue^{-2k\depth}} \right) \frac{\tau/k}{\sqrt{\sur{B} - 2g\eta}} \right] . 
\end{align}

The numerical computation of Lagrangian surface waves is thus done by solving 
the nonlinear expression \rf{eqetaL}, providing a given wave amplitude $\left.
\eta\right|_{t=0} = a$. 
The wave period $T_L$, as well as the Bernoulli constant $B_s$, are both obtained 
from the Lagrangian counterpart of \rf{etamean} and the requirement that 
$\int_{0}^{T_L} \ud \sur{z} = 2\pi\sigma/k$. Moreover, since the abscissa $x(t)$ 
is given explicitly from the definition of the wave profile $\eta(t)$ through 
the real part of formula \rf{zt}, we only have to solve expression \rf{eqetaL} 
for $\eta$ --- and not for both $x$ and $\eta$, as done previously   
by \citet{DSP88} and by \citet{VB94} --- thus allowing us to further reduce the numerical cost.

\section{\label{sec4} Surface recovery from bottom pressure}

Building upon the last three sections, we propose a nonlinear 
integral equation to recover the surface elevation in terms of 
a given measure of pressure $\bot{p}(x)$ at the seabed.
This `measurement' is given here numerically by solving expression 
\rf{eqetaL} and, subsequently, computing the pressure from the 
surface profile (exploiting the boundary integrals and the Bernoulli 
equation).

The principal benefit in establishing an integral formulation 
lies in the fact that no specific eigenfunctions are needed
to fit the pressure data.
In fact, our derived equation remains applicable regardless of 
the nature of the wave under consideration, be it periodic or 
not, overhanging or not. We reemphasise here that, although the 
equations consider $(2\pi/k)$-periodic waves, their aperiodic 
counterparts are easily obtained letting $k\to0^+$. 

\subsection{Eulerian boundary integral formulation}

Let consider the Cauchy integral formula \eqref{dcauchyperegim} 
(without the method of images) for the holomorphic function 
$\Upxi(z) = \mathfrak{P}(z) - g\depth$. It yields
\begin{align}\label{cauchyP}
\mathfrak{P} - g\depth =& \frac{k}{\vartheta} \int_{\mathcal{I}} 
\left[ \plog{0}\!\left(\ue^{\ui k (\sur{z}'-z)}\right) 
(1+\ui\eta_x') (\sur{\mathfrak{P}}' - g\depth) - \plog{0} 
\!\left(\ue^{\ui k (\bot{z}'-z)}\right)(\bot{\mathfrak{P}}' 
- g\depth) \right] \ud x' \nonumber \\
&+ \langle  (1+\ui\eta_x) (\sur{\mathfrak{P}} - g\depth) - 
(\bot{\mathfrak{P}}' - g\depth) \rangle .
\end{align}
In order to simplify the latter expression, we exploit both 
definitions of $\bot{\mathfrak{P}}\eqdef p_b(x)$ and $\ud
\sur{\mathfrak{Q}}/\ud x = (\sur{\mathfrak{P}}-g\depth)(1+
\ui\eta_x)$ (we recall that $\sur{\mathfrak{Q}}$ is a periodic 
function). Hence, expression \rf{cauchyP} can be rewritten 
\begin{align}
\label{cauchyPint}
\mathfrak{P} - g\depth = \frac{\ui}{\vartheta} \int_{\mathcal{I}} 
\frac{\partial}{\partial z} \left[ \plog{1}\!\left( \ue^{\ui k 
(\sur{z}'-z)} \right) (1+\ui\eta_x') (\sur{\mathfrak{P}}' - g\depth) 
-\plog{1}\!\left( \ue^{\ui k (\bot{z}'-z)} \right) (\bot{p}' - g\depth) 
\right] \ud x'.
\end{align}
Before pursuing, let introduce the compact notations for the 
polylogarithmic kernels
\begin{equation}
\mathcal{K}_{\nu} \eqdef \plog{\nu}\!\left( \ue^{\ui k (\sur{z}'-\sur{z})} 
\right) \quad \textrm{and} \quad \mathcal{J}_{\nu} \eqdef \plog{\nu}\! 
\left( \ue^{\ui k (\bot{z}'-\sur{z})} \right) .
\end{equation}
We now evaluate expression \rf{cauchyPint} at the free surface, reducing 
it to
\begin{align}
\label{cauchyPint2}
(\sur{\mathfrak{P}} - g\depth) (1+\ui\eta_x) = \frac{\ui}{2\pi} 
\frac{\ud}{\ud x} \int_{\mathcal{I}} \left[ \mathcal{K}_{1} (1+\ui\eta_x') 
(\sur{\mathfrak{P}}' - g\depth) - \mathcal{J}_{1} (\bot{\mathfrak{P}}' - 
g\depth) \right] \ud x'.
\end{align}
As one can notice, the left-hand side of \rf{cauchyPint2} is the integrand 
of $\sur{\mathfrak{Q}}$. For that reason, we first integrate the whole 
expression over the $x$-coordinate and then split the contributions from 
the surface and the bottom
\begin{align}\label{cauchyQ}
\sur{\mathfrak{Q}} = \frac{\ui}{2\pi} \int_{\mathcal{I}} \mathcal{K}_{1} 
(1+\ui\eta_x') (\sur{\mathfrak{P}}' - g\depth) \ud x' - \frac{\ui}{2\pi} 
\int_{\mathcal{I}} \mathcal{J}_{1} (\bot{p}' - g\depth) \ud x' + K ,
\end{align}
for a constant of integration $K$ also obtained applying condition \rf{etamean}. 

The next step is to decompose the integrand in the first integral of 
\rf{cauchyQ}. To do so, we merely replace the complex velocity by its 
expression \eqref{W} and exploit the identity \rf{intplog} to cancel 
out some constant terms, i.e.
\begin{alignat}{2}
\int_{\mathcal{C}} \mathcal{K}_{1} (\sur{\mathfrak{P}}' - g\depth) 
\ud \sur{z}' &= &&- \frac{1}{2} \int_{\mathcal{C}} \mathcal{K}_{1} 
\sur{W}'^2 \ud \sur{z}' = - \frac{1}{2} \int_{\mathcal{C}} \mathcal{K}_{1} 
\left[ \omega h' + (1-\ui\eta_x') \sqrt{\frac{(\sur{B}-2g\eta')}{(1+\eta_x'^2)}} \right]^2 \ud \sur{z}' \nonumber \\
&= &&-\frac{\omega^2}{2} \int_{\mathcal{C}}\mathcal{K}_{1} h'^2 
\ud \sur{z}' - \omega\sigma \int_{\mathcal{C}} \mathcal{K}_{1} h' 
(1-\ui\eta_x') \sqrt{\frac{(\sur{B}-2g\eta')}{(1+\eta_x'^2)}} 
\ud \sur{z}'  \nonumber \\
& &&-\frac{1}{2} \int_{\mathcal{C}} \mathcal{K}_{1} (1-\ui\eta_x')^2 
\frac{(\sur{B}-2g\eta')}{(1+\eta_x'^2)} \ud \sur{z}' .
\end{alignat}
After some algebraic manipulations, it further reduces to
\begin{alignat}{2}
\label{intsur}
\int_{\mathcal{C}} \mathcal{K}_{1} (\sur{\mathfrak{P}}' - g\depth) 
\ud \sur{z}' &= && -\frac{\omega^2}{2} \int_{\mathcal{C}} \mathcal{K}_{1} \eta'^2 \ud\sur{z}'- \omega^2\depth \int_{\mathcal{C}} \mathcal{K}_{1} \eta' \ud\sur{z}' 
 \nonumber \\
& &&-\frac{1}{2} \int_{\mathcal{I}} \mathcal{K}_{1} (\sur{B} - 2g\eta') (1 - \ui\eta_x') \ud x' \nonumber \\
& &&-\omega\sigma \int_{\mathcal{I}} \mathcal{K}_{1} h' \sqrt{(\sur{B} 
- 2g\eta')(1+\eta_x'^2)} \ud x' \nonumber \\
&= &&-\frac{\ui\omega^2}{k} \int_{\mathcal{I}} \mathcal{K}_{2}
\eta' \eta_x' \ud x' + \frac{\omega^2\depth + g}{k} \int_{\mathcal{I}} \mathcal{K}_{2} \ud x' \nonumber \\
& &&- \int_{\mathcal{I}} \mathcal{K}_{1} (\sur{B} - 2g\eta') \ud x' \nonumber \\
& &&-\omega\sigma \int_{\mathcal{I}} \mathcal{K}_{1} h' \sqrt{(\sur{B} 
- 2g\eta')(1+\eta_x'^2)} \ud x' .
\end{alignat}

Substituting \eqref{intsur} into \eqref{cauchyQ}, the imaginary part 
yields the  Eulerian integral formulation for the surface recovery 
\begin{align} \label{intpeta}
2\pi \Im \{ \sur{\mathfrak{Q}} \} = &\frac{\omega^2}{2k} \int_{\mathcal{I}} 
\Im \{ \mathcal{K}_{2} \} (\eta'^2)_x \ud x' - \int_{\mathcal{I}} 
\Re \{ \mathcal{K}_{1} \} (\sur{B} - 2g\eta') \ud x'  \nonumber \\
&+\frac{\omega^2\depth + g}{k} \int_{\mathcal{I}} \Re \{ \mathcal{K}_{2} \} 
\ud x'- \int_{\mathcal{I}} \Re \{ \mathcal{J}_{1} \} (\bot{\mathfrak{P}}' - g\depth) \ud x' \nonumber \\
& - \omega\sigma \int_{\mathcal{I}} \Re \{ \mathcal{K}_{1} \} 
h' \sqrt{(\sur{B} - 2g\eta')(1+\eta_x'^2)} \ud x' - 2\pi\Im\{K\},
\end{align}
where $K$ is the same constant as in \rf{cauchyQ}.

Equation \eqref{intpeta} is a nonlinear integral equations for the 
free surface recovery from the bottom pressure. Being strictly Eulerian, 
this equation is not suitable for overhanging waves. For the latter, one 
can proceed as follow.

\subsection{Hybrid formulation}

Equation \eqref{intpeta} involves integrals at the free surface and at the bottom. 
In practice, the bottom pressure is given at some known abscissa $x$, so the bottom 
integral must be kept in Eulerian form. However,  Eulerian integrals are not suitable 
for overhanging waves, so we rewrite surface integrals in their Lagrangian counterparts.   
Doing so, the inverse problem is described by a mixed 
Eulerian--Lagrangian formalism.

Thus, the surface integrals in \eqref{intpeta} being rewritten in Lagrangian description, 
the bottom integral being kept in Eulerian form, one gets the general expression for 
the surface recovery
\begin{align} \label{intpetalag}
2\pi \Im \{ \sur{\mathfrak{Q}} \} = &\frac{\omega^2}{k} \int_{0}^{T_L} \left[ k^{-1} 
\Re \{ \mathcal{K}_{3} \} + \left( h' + g\omega^{-2} \right) \Re \{ \mathcal{K}_{2} \}  
\right] \cos(\sur{\theta}') \sqrt{\sur{B} - 2g\eta'} \ud t'      \nonumber \\
&- \int_{0}^{T_L} \Re \{ \mathcal{K}_{1} \} \left[ \cos(\sur{\theta}') \sqrt{\sur{B} 
- 2g\eta'} + \sigma\omega h' \right]\left(\sur{B} - 2g\eta'\right)\ud t' \nonumber \\
&- \int_{0}^L \Re \{ \mathcal{J}_{1} \} (\bot{p}' - g\depth) \ud x' - 2\pi\Im\{K\} .
\end{align}

This reformulation is necessary for practical recovery of overhanging waves. 
It is of course also suitable for non-overhanging waves.

\section{\label{sec5} Numerical illustrations}

\begin{figure}[t!]
    \centering
    \subfloat[]{
    \includegraphics*[width=.95\textwidth]{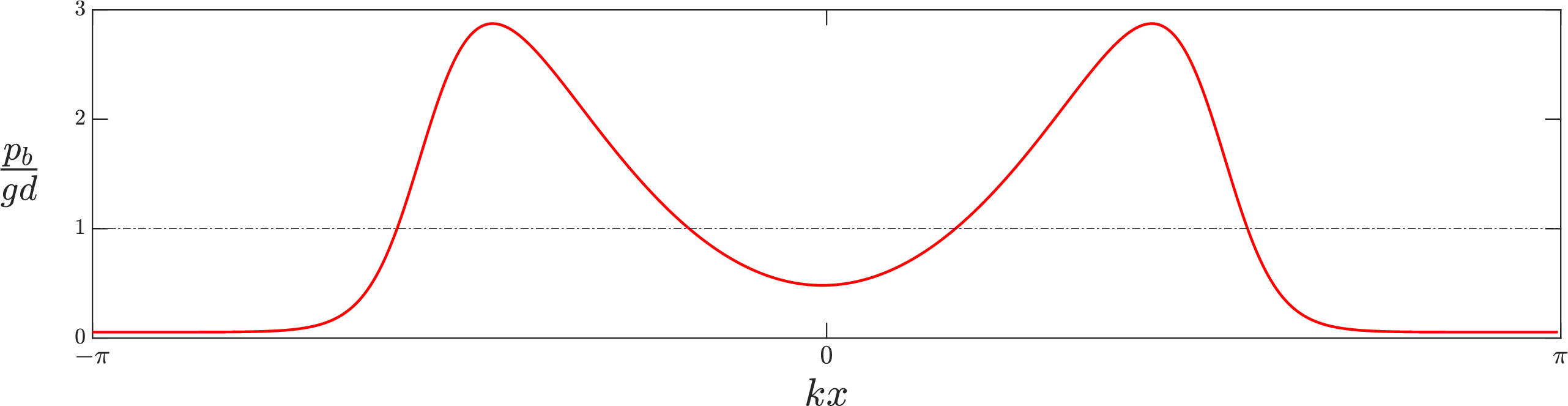} \label{Fig2a}
    } \\
    \subfloat[]{
    \hspace{-.1em}
    \includegraphics*[width=.95\textwidth]{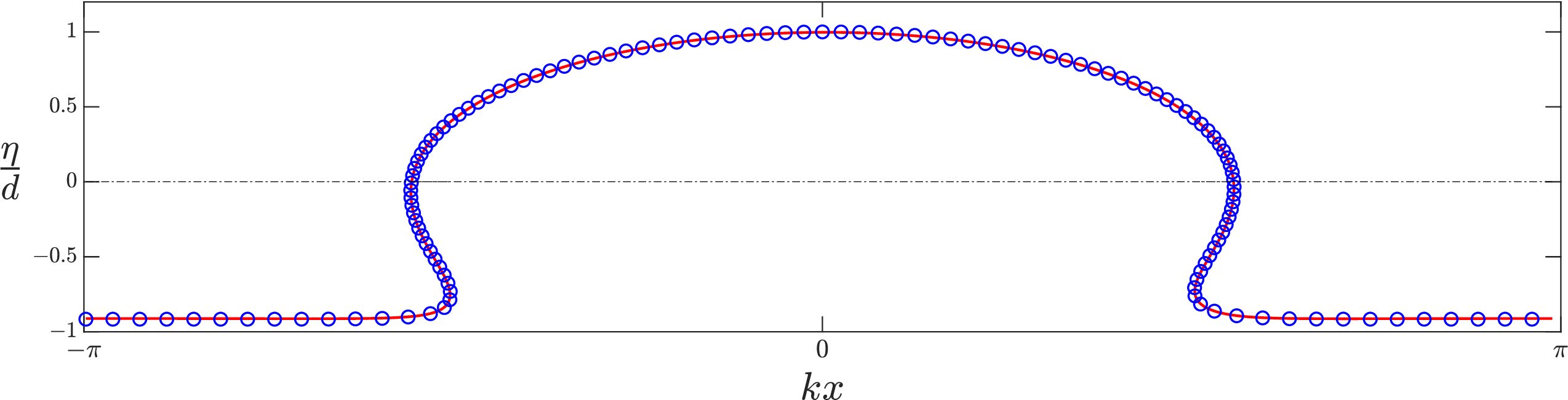} \label{Fig2b}
    }
    \caption{Numerical demonstration of the surface recovery procedure for an overhanging and periodic steady wave with a constant vorticity $\omega\sqrt{d/g}=3\sqrt{2}$.
    (a) Eulerian representation of the bottom pressure. The dashed dot line corresponds to the mean bottom pressure.
    (b) Surface wave profile (blue circles) obtained from expression \rf{eqeta}. The red line represents the surface reconstruction achieved from $\bot{p}$ through equation \rf{intpetalag}, while the dashed dot line indicates the mean water level.}
    \label{Fig2}
\end{figure}

\subsection{Details on the overall methodology}

From a numerical standpoint, the nonlinear integral equations \eqref{eqeta} 
and \eqref{intpetalag}, employed respectively for computing the surface profile 
and recovering it from the bottom pressure, possess notable characteristics. 
First, these equations eliminate the need for evaluating the derivative of 
$\sur{\theta}$ at any point. Second, we can rely on Fourier analysis since the kernels
are periodic and use trapezoidal rule for numerical integration.

Since we already know the value of $g$ and the wavenumber $k$ (or equivalently 
the wave period $L$) as inputs in our numerical scheme, we can deduce the depth 
of the layer $d$ from the definition of the hydrostatic law 
\citep{Clamond2013,CLH23}. For simplicity, 
we consider that the constant vorticity $\omega$ is known. 
However, $\omega$ can also be determined adapting the procedure 
described by \citet{CLH23}.
Then, initialising our numerical schemes with an appropriate 
initial guess, either from linear theory or by a previous 
iteration, as done by \citet{DSP88}, we 
observe fast convergence for a discrete set of $N$ equidistant points. 
In our simulations, we typically use $N=128$ for moderate waves and $N=512$ 
for waves (regardless of whether they exhibit overturning or not) with large 
amplitudes. 

When investigating overhanging waves, the non-algebraic nonlinearities inherent 
in the problem often pose numerical challenges. The most troublesome issue arises 
from aliasing errors present in the functions spectra, a phenomenon occasionally 
referred to as ``spectral blocking'' \citep{Boyd2001}, leading to exponential 
growth of high frequencies. 
As a consequence, this aliasing effect prevent the spectral accuracy 
inherent in our formulation. To address this issue, we employ the ``zero padding'' 
method, which involves increasing the size of the quadrature in Fourier space while 
appending zeros above the Nyquist frequency. 
Subsequently, we transform the functions back to physical space, compute the 
nonlinear terms, and filter out the previously introduced zero frequencies from 
the spectrum. 
For quadratic nonlinear terms, enlarging the degree of quadrature by a factor of 
$3/2$ has proven sufficient to mitigate this phenomenon \citep{Orszag1971}. 
However, in the context of non-algebraic nonlinearities encountered in this study, 
this argument does not hold, and the exact value of the enlargement factor remains 
unknown. 
Instead, we utilize a factor of $2$ (typically suitable for cubic nonlinearities) 
when performing products throughout the algorithm. 
Although we experimented with larger factors in our simulations, it appeared that 
this value was adequate for eliminating spurious frequencies in most aliased spectra.

In addition to aliasing, we noticed that it is numerically more efficient
to perform a change of coordinates in evaluating the finite integrals.
Particularly, significant variations in $\sur{\theta}$ occur where the wave 
undergoes overturning, indicating a requirement for additional collocation points 
in these regions. 
Thus, rather than evaluating the previous integrals with respect to the time 
variable $t\in[0,T_L]$, we introduce a new integration variable $\Upxi(t)$, defined 
as
\begin{equation}
\label{xi}
\frac{\ud \xi}{\ud t} \eqdef \sqrt{1 + \beta\sin\!\left(\sur{\theta}\right)^2 } ,
\end{equation}
where $\beta$ is a scaling parameter arbitrarily set. 
A similar change of coordinate can be found in \citep{DSP88}.
In most simulations involving overhanging waves, we employ a value of 
$\beta=2\pi/(kdN)$. 

Finally, we solve the whole system of equations \eqref{eqeta} 
and \eqref{intpetalag} with the built-in iterative solver 
\textsf{fsolve} from \textsc{Matlab} software, using the 
Levenberg--Marquardt algorithm.

\begin{figure}[t!]
    \centering
    \subfloat[]{
    \includegraphics*[width=.49\textwidth]{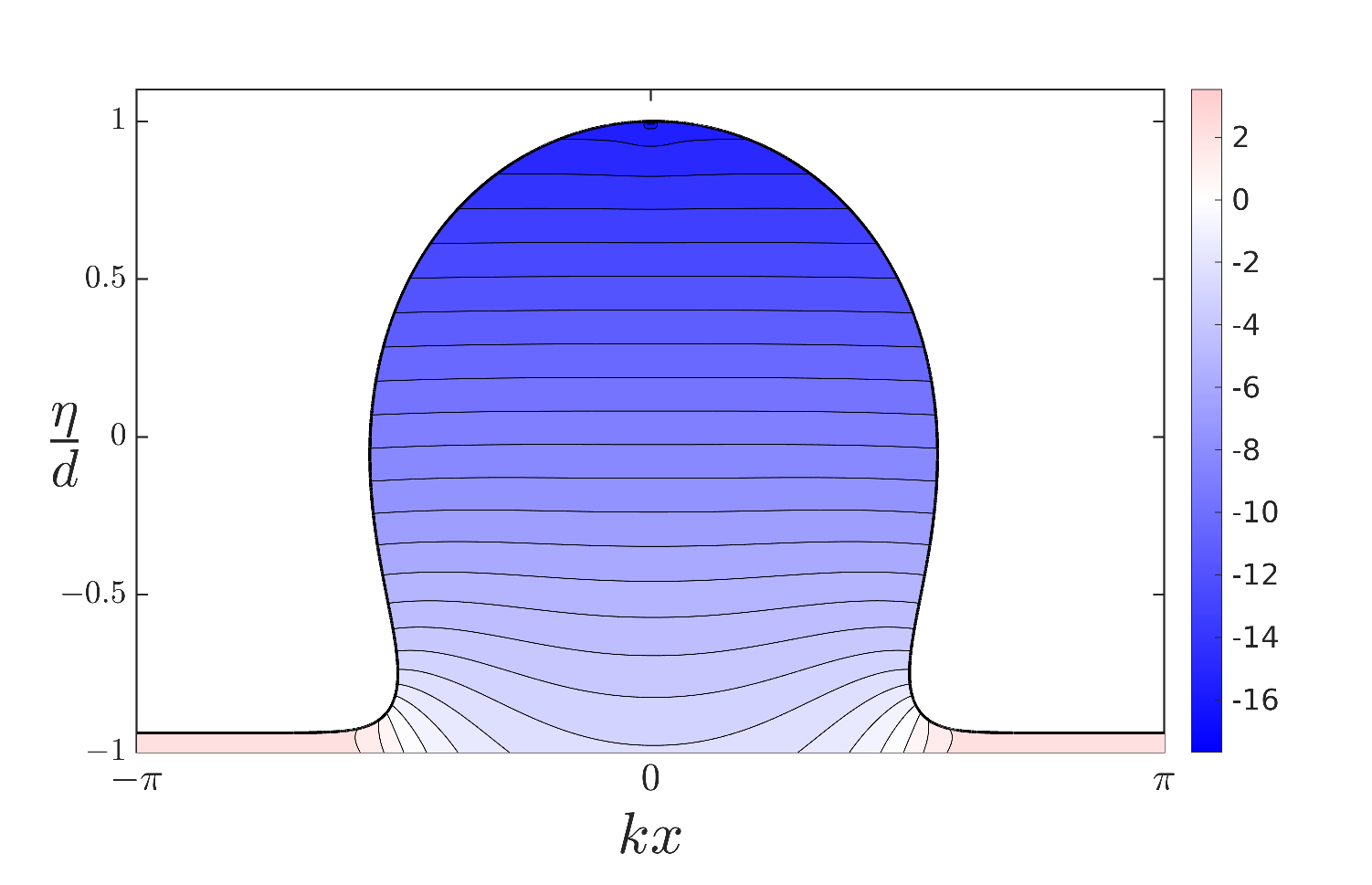} 
    \label{Fig3a}}
    \subfloat[]{
    \includegraphics*[width=.49\textwidth]{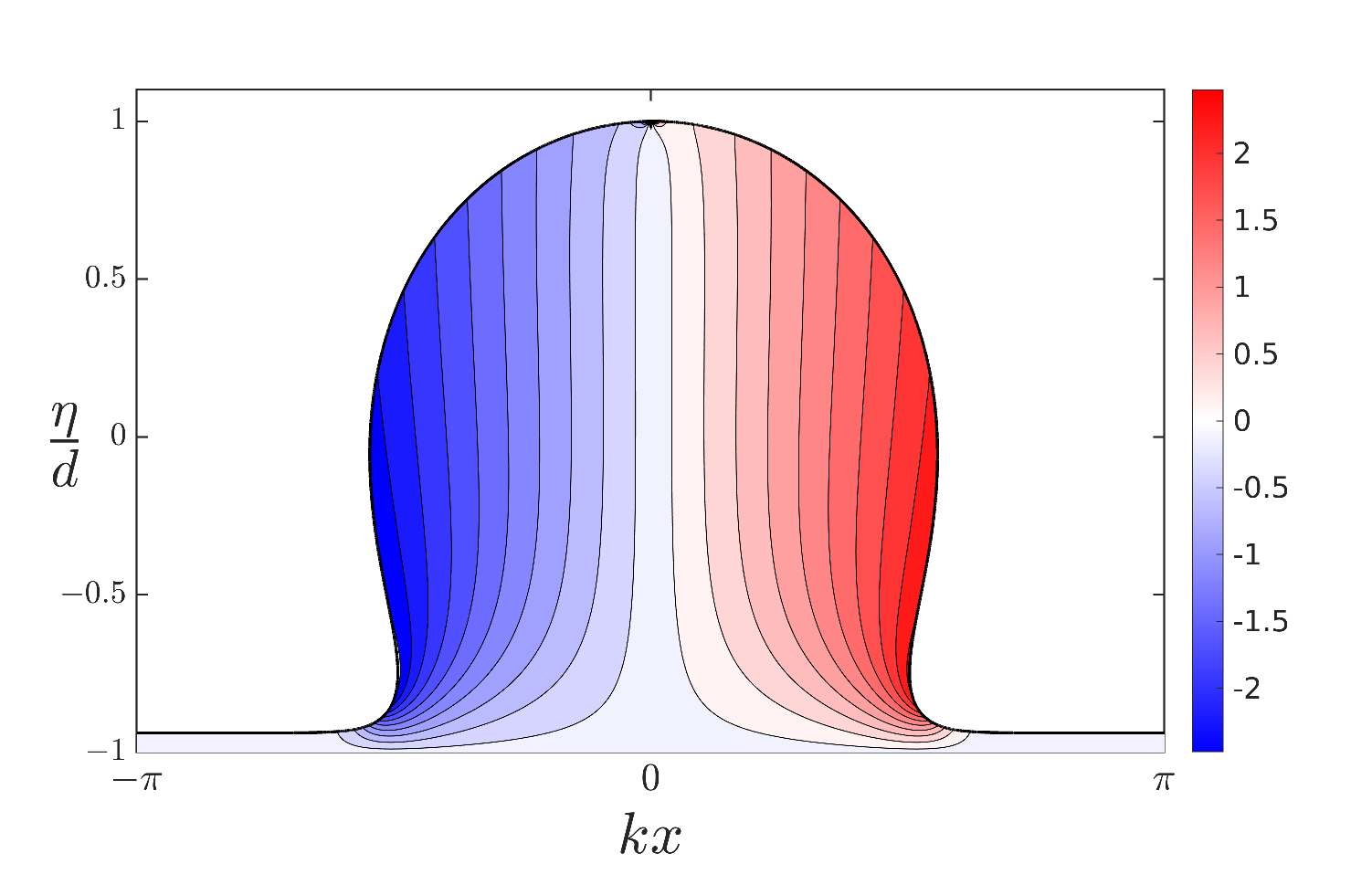} 
    \label{Fig3b}}
    \vspace{-1em}\\
    \subfloat[]{
    \includegraphics*[width=.49\textwidth]{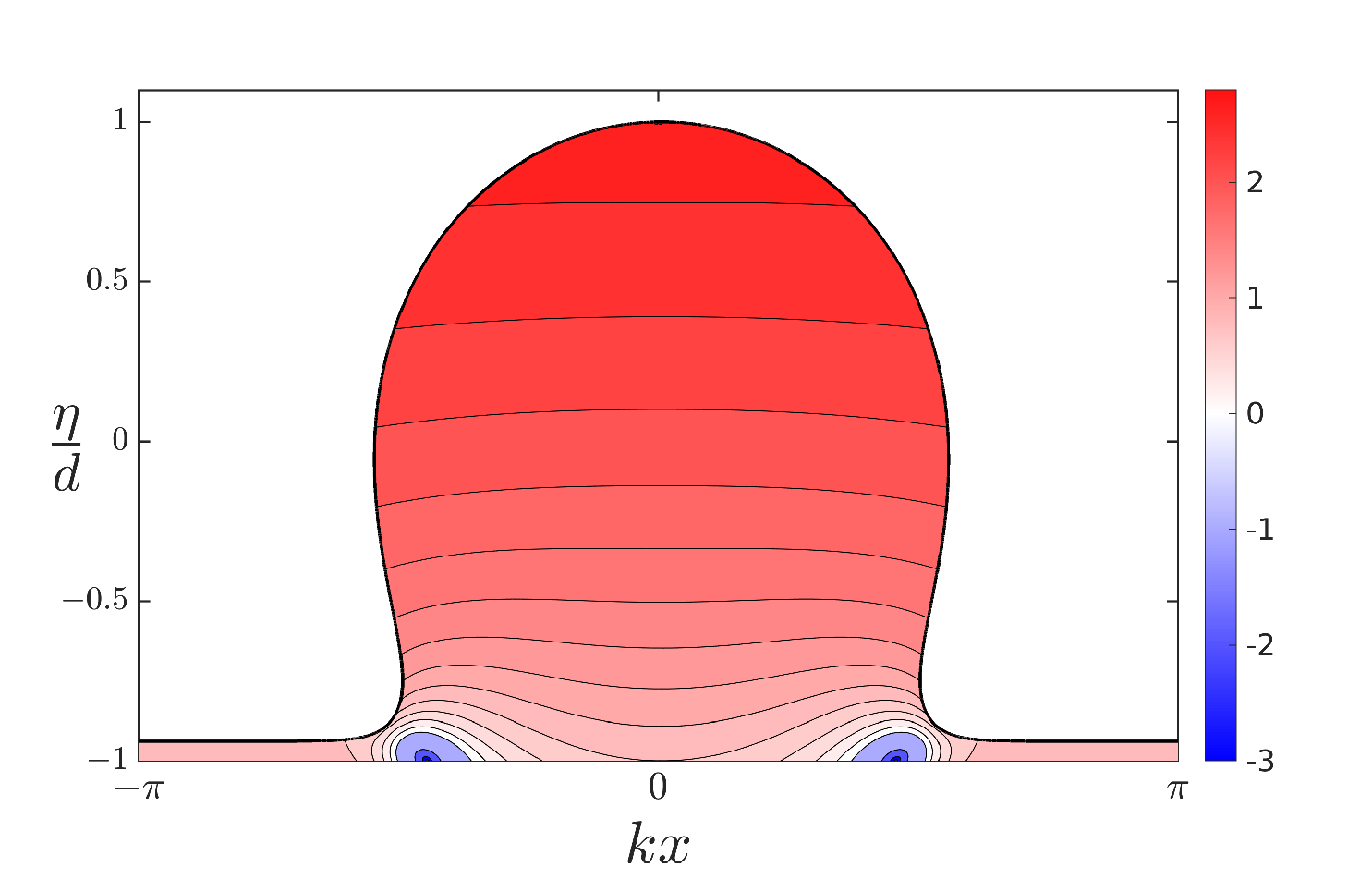} 
    \label{Fig3c}}
    \subfloat[]{
    \includegraphics*[width=.49\textwidth]{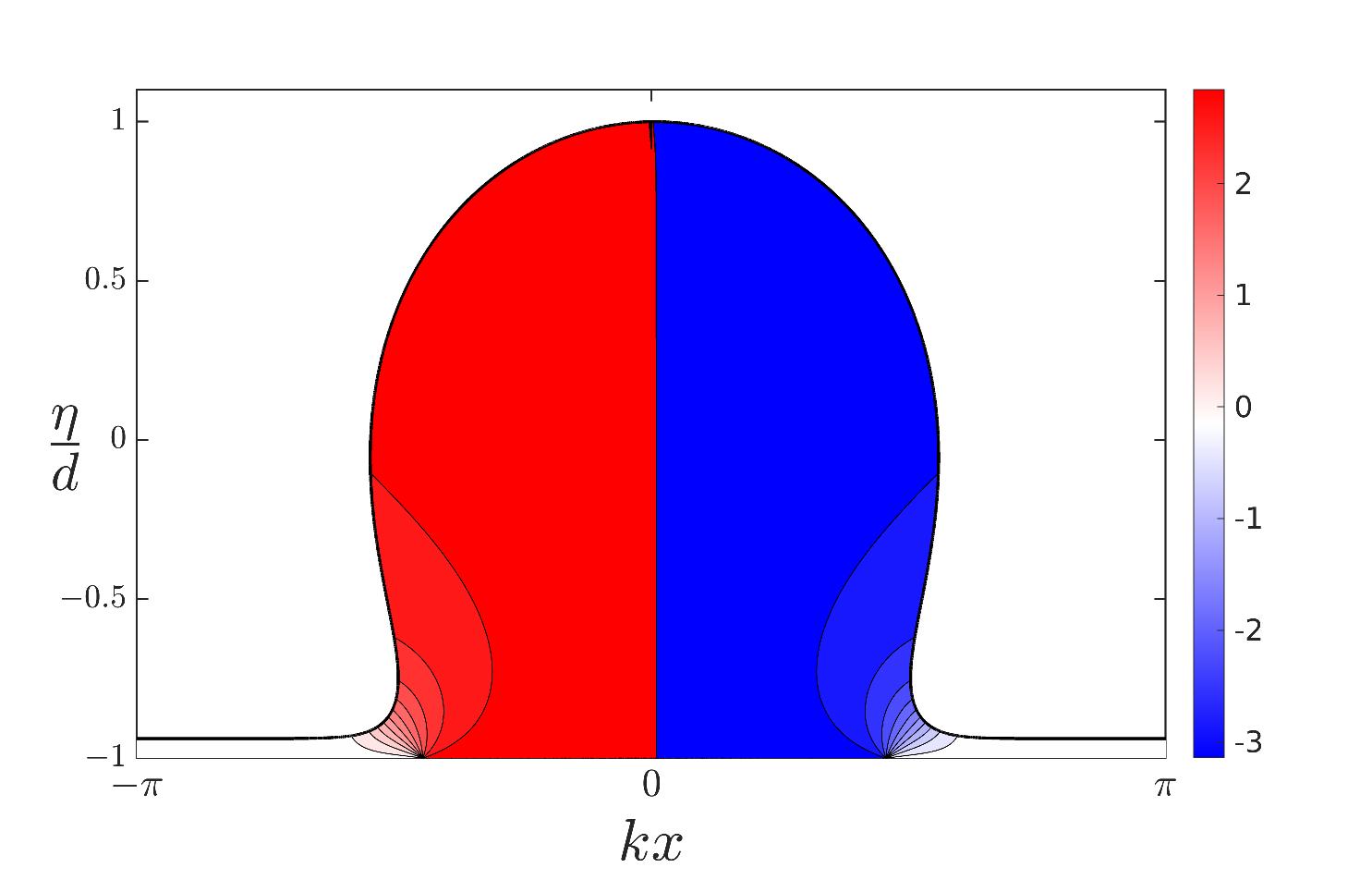} 
    \label{Fig3d}}
    \caption{Iso-values of (a): $u/\sqrt{gd}$, (b): $v/\sqrt{gd}$, (c): $\log{(\vert w\vert/\sqrt{gd})}$ and (d): $\arg{( w/\sqrt{gd})}$ for the  configuration displayed in figure \eqref{Fig2}.}
    \label{Fig3}
\end{figure}

\subsection{Periodic overhanging wave}

The first case of interest, as shown in figure \ref{Fig2}, depicts a periodic 
overturning wave of large amplitude. 
This particular scenario poses significant computational challenges, making it 
an ideal benchmark for evaluating the robustness of our recovery procedure.
Utilizing the pressure data highlighted in figure \ref{Fig2a}, we successfully 
reconstruct the surface profile using the expression \ref{intpetalag}, resulting 
in excellent agreement, as evident from figure \ref{Fig2b}. 
Notably, the change of variables achieved through equation \ref{xi} effectively 
concentrates the points in regions where $\theta_s$ varies the most. 
Consequently, we attain a high level of accuracy, with $\vert\vert\eta - 
\eta^{\textrm{ex}}\vert\vert_{\infty} \approx 4.48\times10^{-3}$, where 
$\eta^{\textrm{ex}}$ corresponds to a numerical solution obtained by 
solving equation \eqref{eqetaL}. 
Furthermore, for the computation of the unknown Bernoulli constant at the 
surface, the error is approximately $\vert\sur{B} - \sur{B}^{\textrm{ex}}
\vert \approx 5.83\times10^{-3}$.


We emphasize the effectiveness of our recovery procedure by 
presenting the velocity field within the fluid layer in both upper panels of 
figure \ref{Fig3}. 
Notably, a pair of stagnation points is observed in the panel \ref{Fig3c} on the flat bottom boundary, 
which often poses challenges when utilizing conformal mapping techniques. 
In the present context, since our methodology operates solely in the physical 
plane, our general approach can readily reconstruct the surface profile 
regardless of the presence or absence of stagnation points.

\begin{figure}[t!]
    \centering
    \subfloat[]{
    \includegraphics*[width=.955\textwidth]{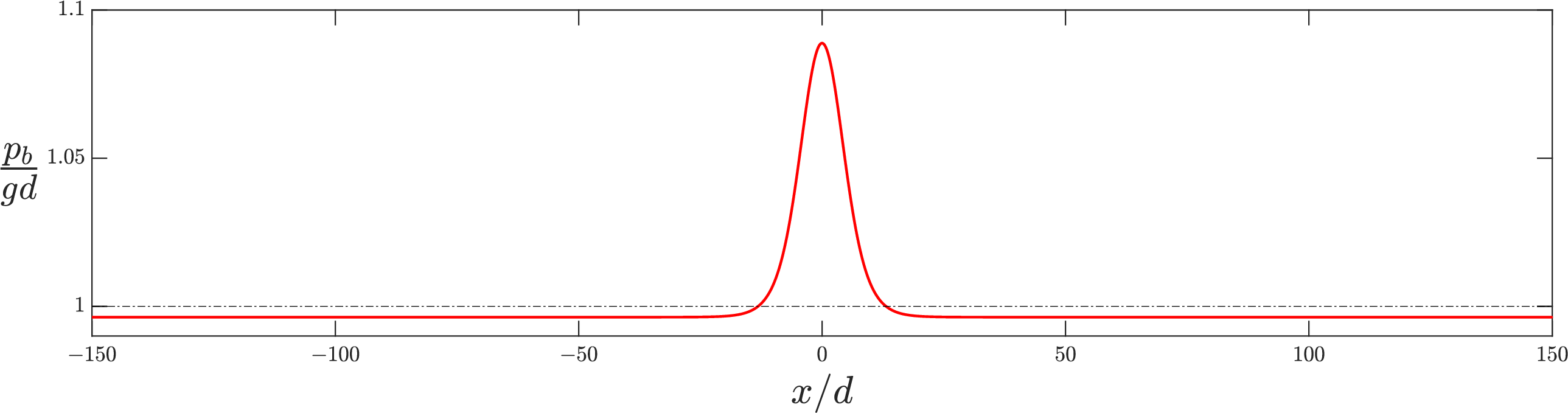} \label{Fig4a}
    } \\
    \subfloat[]{
    \hspace{-.1em}
    \includegraphics*[width=.95\textwidth]{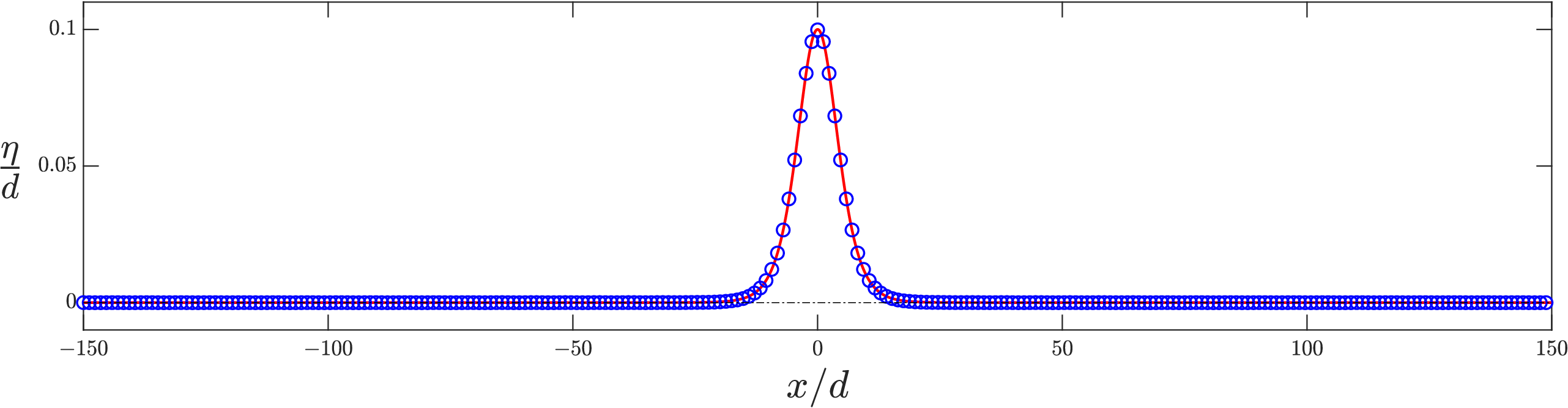} \label{Fig4b}
    }
    \caption{Surface recovery procedure for a solitary steady wave with a constant vorticity $\omega\sqrt{d/g}=1$. Both panels have the same legend than figure \ref{Fig2}.}
    \label{Fig4}
\end{figure}

\subsection{Rotational solitary surface wave}

Computing solitary waves using our procedure is straightforward but 
necessitates considering a relatively large numerical domain to accurately 
capture their behavior far from the crest. 
Meanwhile, to ensure that the wave is located above the mean water level, 
we substitute the condition \eqref{etamean} with 
\begin{equation}
\eta^{(n+1)} = \eta^{(n)} - \min\eta^{(n)} ,
\end{equation}
instead, at each numerical iteration $n$.

The recovery process for a solitary wave is presented in figure \ref{Fig4}, 
which illustrates the pressure data in the upper panel and the corresponding 
surface profile in the lower panel. 
We consider here a solitary wave of relatively small amplitude because 
it is quite challenging. Indeed, the larger the solitary wave, the 
faster its decay (thus requiring a smaller computational box) and 
the larger the ratio signal/noise (for field data). So, in that respect, 
the recovery of small amplitude solitary waves is more challenging. 

In the present specific case, the agreement with a given numerical solution 
is great, yielding numerical errors of $\vert\vert\eta - \eta^{\textrm{ex}}
\vert\vert_{\infty} \approx 1.46\times10^{-4}$ and $\vert\sur{B} - 
\sur{B}^{\textrm{ex}}\vert \approx 4.21\times10^{-5}$ for the surface profile 
and the Bernoulli constant, respectively.
Unfortunately, spurious infinitesimal oscillations (located) far from the wave 
crest) prevent us to reach a much better accuracy that we would expect from 
our method.
In order to facilitate these computations and remove unwanted oscillations, 
another change of variables can be implemented (similar to the approach 
\eqref{xi} used for the periodic case) to concentrate the quadrature points 
near the crest, rather than far away where the elevation is infinitesimal. 
However, we reserve this task for future investigations, which will provide 
more comprehensive details on the efficient computation of solitary waves 
within this context.

\section{Discussion}\label{sec6} 

This work presents a novel and comprehensive boundary integral 
method for recovering surface water-waves from bottom pressure 
measurements. Despite the inherent complexity of this inverse 
problem, we successfully formulate the relatively simple expression 
\eqref{intpetalag} for surface recovery, enabling the computation 
of a wide range of rotational steady waves. A significant advantage 
of this approach lies in the integral formulation, which eliminates 
the need for arbitrarily selecting a basis of functions to fit the 
pressure data, as done previously in 
\citep{Clamond2013,ClamondConstantin2013,CLH23}.

To demonstrate the robustness and efficiency of our method, we 
showcased two challenging examples: an overturning wave with a 
large amplitude and a solitary wave. In both cases, we accurately 
recovered the surface profile and the hydrodynamic parameters with 
good agreement. Although it might be possible to adapt our numerical 
procedure to compute extreme waves (with angular surface) with (or 
without) overhanging profiles, this task is left for future 
investigations. In fact, our main goal here is a proof of concept 
and to provide clear evidence on the effectiveness of this new 
formulation.

In conclusion, this article, along with the proposed boundary 
integral formulation, represents a significant milestone in solving 
the surface wave recovery problem, providing a solid foundation 
for future extensions, such as its potential application to 
three-dimensional configurations. Indeed, in 3D holomorphic 
functions cannot be employed but integral representations via 
Green functions remain, so an efficient fully nonlinear surface 
recovery is conceivable.  
\\

\noindent{\bf Funding.} Joris Labarbe has been supported by the 
French government, through the $\mbox{UCA}^{\mbox{\tiny JEDI}}$ 
{\em Investments in the Future\/} project managed by the National 
Research Agency (ANR) with the reference number ANR-15-IDEX-01. \\

\noindent{\bf Declaration of interests.} The authors report no 
conflict of interest.

\appendix

\section{\label{appA} Logarithms and Polylogarithms}

The function $\ln(x)$ denotes the {\em natural logarithm\/} 
(Napierian logarithm) of a real positive variable $x$ 
($x\in\mathds{R}^+$), and $\log(z)$
denotes the {\em principal logarithm\/} of a complex variable
$z\in\mathds{C}$, i.e.,
\begin{equation} \label{deflog}
\log(z) \eqdef \ln|z| + \ui\arg(z), \qquad - \pi<\arg(z)\leqslant\pi.
\end{equation}
This definition requires that the argument of any complex 
number lies in $]-\pi;\pi]$. It implies, in particular, that 
$\arg(z^{-1})=-\arg(z)$ if $z\not\in\mathds{R}^-$ and that 
$\arg(z^{-1})=\arg(z)=\upi$ if
$z\in\mathds{R}^-$. We have the special relations
\begin{align}
\log(-z) &= \ui\upi + \log(z) + 2\ui\upi
\left\lfloor-\arg(z)/2\upi\right\rfloor, \\
\log\!\left(\ue^{\ui\/z}\right) &= \ui z + 2\ui\upi
\left\lfloor\half - \Re(z/2\upi)\right\rfloor, \\
\log\!\left(-\ue^{\ui\/z}\right) &= \ui (\upi+z) + 2\ui\upi\left\lfloor - \Re(z/2\upi)\right\rfloor,
\end{align}
where $\left\lfloor\cdot\right\rfloor$ is the rounding toward $-\infty$.

The polylogarithms can be defined, for $|z|<1$ and $\nu\in\mathds{C}$, by
\begin{align} \label{defnulog}
\plog{\nu}{(z)} &= \sum_{n=1}^\infty{\frac{z^n}{n^\nu}} ,
\end{align}
and for all complex $z$ by analytic continuation \citep{Wood1992}. 
With the above definition of the complex logarithm, we have the special 
inversion formulae
\begin{align} 
\plog{0}(z) + \plog{0}(z^{-1}) + 1 &= 0, \label{nulloginv} \\
\plog{1}(z) - \plog{1}(z^{-1}) + \log(-z) &=
\left\{\begin{array}{lrl}
    0  \qquad     & \qquad & \text{if}\quad z\not\in[0;1], \\
    2\/\ui\/\pi   & \qquad & \text{if}\quad z\in[0;1],
\end{array}\right. \\
\plog{2}(z) + \plog{2}(z^{-1}) + \frac{1}{2} \log^2(-z) + \frac{1}{6} \upi^2 &=
\left\{\begin{array}{lrl}
    0  \qquad           & \ & \text{if}\quad z\not\in[0;1], \\
    2\/\ui\/\pi\log(z)  & \ & \text{if}\quad z\in[0;1],
\end{array}\right. \label{diloginv}
\end{align}
The polylogarithms for $\nu\in\mathds{N}^*$ are single-valued functions in the cut plane
$z\in\mathds{C}\backslash [1;+\infty[$ and the inversion formula can be used to extend 
their definition for $z\in[1;+\infty[$. Note that the inversion formula depends on the 
definition of the principal logarithm that is not unique, thus several variants can be 
found in the literature. 

We finally note other relations useful in this paper
\begin{align}
\plog{\nu}\!\left(\ue^{\pm\/\ui\/z}\right) &= \mp \ui\ud
\plog{\nu+1}\!\left(\ue^{\pm\/\ui\/z}\right)/\ud\/z, \label{Lidz} \\
\plog{0}\!\left(\ue^{\pm\/\ui\/z}\right) &= \pm\/\ihalf\/\cot\!\left(z/2\right) 
- \half , \label{Licot} \\
\plog{1}\!\left(\ue^{\pm\ui z}\right) &= \pm\/\ui\upi\left[\Re\{z\}/2\upi\right]
+ \ihalf\arg(z^2) - \ui\arg(\mp\ui z) 
- \half\/\log\!\left(4\sin^2(z/2)\right) \mp \ihalf z \\
\plog{2}\!\left(\ue^{\pm\/\ui\/z}\right) &= \frac{\upi^2}{6} + \frac{z^2}{4} \mp 
\left(\arg(z^2)-2\arg(\mp\ui z)\right)\frac{z}{2} 
\mp \frac{\ui}{2}\int_0^z\log\!\left(4\sin^2\!\left(\frac{z'}{2}\right)\right)\/\ud\/z', 
\end{align}
where $[\cdot]$  denotes the rounding toward zero, the last relation being valid for 
$-2\upi<\mathrm{Re}\{z\}<2\upi$.

\section{\label{appB} Singular Leibniz integral rule}

Let be $K(x,y)$ a function regular everywhere for $y\in\interval{a}{b}$, except perhaps at 
$y=y_0\in\interval[open]{a}{b}$ ($y_0$ generally depending on $x$, as well as $a$ and $b$) 
where $K$ may be singular, its finite integral being taken in the sense of 
Cauchy principal value, i.e. 
\begin{equation}
J(x) = \dashint_{a(x)}^{b(x)} K(x,y) \ud y \eqdef \lim_{\epsilon\to0^+} \left\{
 \int_{a(x)}^{y_0(x)-\epsilon} K(x,y) \ud y + \int_{y_0(x)+\epsilon}^{b(x)}
K(x,y) \ud y \right\} ,
\end{equation}
exists. The first derivative of $J$ is thus
\begin{align}
\label{reldJ}
\frac{\ud J}{\ud x} &= \lim_{\epsilon\to0^+} \left\{ \frac{\ud}{\ud x}
\int_{a}^{y_0-\epsilon} K(x,y) \ud y + \frac{\ud}{\ud x} \int_{y_0+\epsilon}^{b}
K(x,y) \ud y \right\} \nonumber \\
&= \lim_{\epsilon\to0^+} \left\{ \int_{a}^{y_0-\epsilon} \frac{\partial K(x,y)}{\partial x}
 \ud y + \frac{\ud y_0}{\ud x} K(x,y_0-\epsilon) - \frac{\ud a}{\ud\/x}
K(x,a) \right. \nonumber \\
&\qquad\qquad\left. + \int_{y_0+\epsilon}^{b} \frac{\partial K(x,y)}{\partial x}
\ud y + \frac{\ud b}{\ud x} K(x,b) - \frac{\ud y_0}{\ud x} K(x,y_0+\epsilon) \right\} \nonumber \\
&= \dashint_{a}^{b} \frac{\partial K(x,y)}{\partial x} \ud y + \frac{\ud b}
{\ud x} K(x,b) - \frac{\ud a}{\ud x} K(x,a) \nonumber \\
&\quad - \frac{\ud y_0}{\ud x} \lim_{\epsilon\to0^+} \left\{ K(x,y_0+\epsilon) - K(x,y_0-\epsilon) \right\} . 
\end{align}

For instance, $a$ and $b$ being constant and for a sufficiently well-behaving function 
$\varphi$, we have
\begin{align}
\dashint_{a}^{b} \frac{\varphi(y)}{x-y} \ud\/y &= \dashint_{a}^{b} \frac{\partial
\ln|x-y| \varphi(y)}{\partial\/x} \ud\/y \nonumber \\ 
&= \frac{\ud}{\ud\/x}\dashint_{a}^{b} \ln|x-y| \varphi(y)\ud\/y + \lim_{\epsilon\to0^+} \ln(\epsilon) \left[ \varphi(x+\epsilon) - \varphi(x-\epsilon) \right] ,
\label{relcauchdev1}
\end{align}
where the limit is zero if $\varphi$ is H\"older continuous (sufficient but 
unnecessary condition).
This formula is a consequence of \eqref{reldJ} but also of the choice of 
the antiderivative of $1/(x-y)$. Indeed, one can also write  
\begin{align}
\dashint_{a}^{b} \frac{\varphi(y)}{x-y} \ud\/y &= \dashint_{a}^{b} \frac{\partial
\log(x-y) \varphi(y)}{\partial\/x} \ud\/y \nonumber \\
&= \frac{\ud}{\ud\/x}\dashint_{a}^{b} \log(x-y) \varphi(y) \ud\/y + \ui\pi\varphi(x) + \lim_{\epsilon\to0^+} \ln(\epsilon) \left[ \varphi(x+\epsilon) - \varphi(x-\epsilon)
\right] ,
\label{relcauchdev2}
\end{align}
where we have used the relations $\log(\epsilon)=\ln(\epsilon)$ and $\log(-\epsilon)=
\ln(\epsilon)+\ui\pi$, since $\epsilon\in\mathds{R}^+$. 
The relation \eqref{relcauchdev1} is more convenient when dealing only with real variables, while \eqref{relcauchdev2} is more suitable for complex formulations.
The reason behind the latter argument is because $\log(x)$ can be continued analytically 
in the complex plane, which is not the case with $\ln|x|$. 


\addcontentsline{toc}{section}{References}
\bibliographystyle{abbrvnat}
\bibliography{Biblio}

\end{document}